\documentstyle[12pt]{article}

  \newtheorem{theo}{Theorem}
  
  \newtheorem{prop}{Proposition}
  
  \newtheorem{lemma}{Lemma}

  \newcommand{\be}{\begin{equation}}
  \newcommand{\ee}{\end{equation}}
  \newcommand{\bea}{\begin{eqnarray}}
  \newcommand{\eea}{\end{eqnarray}}
  \newcommand{\lb}{\label}
\renewcommand{\a}{\alpha}
  
\renewcommand{\b}{\beta}
  
  \newcommand{\g}{\gamma}

  \newcommand{\e}{\mbox{\rm e}}
  
\renewcommand{\O}{\Omega}

\renewcommand{\l}{\lambda}
\renewcommand{\L}{\Lambda}
  \newcommand{\s}{\sigma}
\renewcommand{\S}{\Sigma}

  \newcommand{\bcal}[1]{{\cal #1}}
  \newcommand{\cala}{\bcal{A}}
  
  \newcommand{\calc}{\bcal{C}}

  \newcommand{\Ad}{\mbox{\rm Ad}}

  \newcommand{\ldel}{\langle}
  \newcommand{\rdel}{\rangle}
  \newcommand{\sprod}    [2]{\ldel \,#1\,,\,#2\,\rdel}

  \newcommand{\lbrac}[2]{[#1,#2]}

  \newcommand{\st} [1]{{\mbox{\scriptsize #1}}}

  \newcommand{\sthree}{\mbox{\rm S}^3}
  
  \newcommand{\rthree}{\mbox{\rm\bf R}^3}
  \newcommand{\cdirac}  {{\cal D}}
  \newcommand{\tcdirac} {\tilde{\cdirac}}
  \newcommand{\bdirac}  {\mbox{\bf D}}
\renewcommand{\vec}  [1]{\mbox{\underline #1}}

  \newcommand{\ind}{\mbox{\rm ind}}
  \newcommand{\ctr}     {{\mbox{\scriptsize C}}}
  
  \newcommand{\cptr}     {{\mbox{\scriptsize CP}}}
  \newcommand{\tcptr}     {{\mbox{\scriptsize\bf cp}}}
  \newcommand{\compl}   {{\mbox{\scriptsize\bf C}}}
  \newcommand{\bbb}  [1]{\mbox{\bf #1}}

  \newcommand{\ncs}{N_{\st{CS}}}

\begin{document}

\begin{flushright}
Zurich University Preprint\\
ZU-TH 3/97
\vskip 0.5\baselineskip
hep-th/9702119
\end{flushright}

\renewcommand{\theequation}{\arabic{section}.\arabic{equation}}
\setcounter{equation}{0}

\thispagestyle{empty}

\vfill
\begin{center}
{$\!\!\!${\Large\bf
\noindent \hbox{Zero Modes of the Dirac Operator for regular}

\medskip
 Einstein-Yang-Mills Background fields
}}
\end{center}
\vspace{5 ex}
\vfill
\begin{center}
{\bf Othmar Brodbeck and Norbert Straumann}
\vskip 0.5cm
Institute for Theoretical Physics\\University of Zurich\\
Winterthurerstrasse 190, CH-8057 Zurich
\end{center}
\vfill
\begin{quote}
The existence of normalizable zero modes of the twisted Dirac operator
is proven for a class of static Einstein-Yang-Mills background fields
with a half-integer Chern-Simons number. The proof holds for any
gauge group and applies to Dirac spinors in an arbitrary representation
of the gauge group. The class of
background fields contains all regular, asymptotically
flat, CP-symmetric
configurations with a connection that is globally described by a
time-independent spatial one-form which vanishes sufficiently fast at
infinity. A subset is provided by all neutral, spherically symmetric
configurations which satisfy a certain genericity condition, and for
which the gauge potential is purely magnetic with real magnetic
amplitudes.
\end{quote}
\vfill
\section{Introduction}
A few years ago, Gibbons and Steif \cite{gibbons} studied the Dirac
equation in the background of the static, particle-like
Bartnik-McKinnon
solution \cite{bartnik} of the Einstein-Yang-Mills (EYM) theory. By an
explicit calculation they found a normalizable zero-energy state in the
\hbox{S-wave} sector and speculated on its possible role in anomalous
fermion production. There is indeed a close analogy with the situation
in the electroweak theory which has been investigated by many authors.
(For a recent review see
\cite{rubakov}.) As was first pointed out by Gal'tsov and Volkov
\cite{MVsphaleron}, the Bartnik-McKinnon solution can be interpreted as
a sphaleron, in other words, as a static saddle point solution
separating vacua with different Chern-Simons number. One type of
instabilities is related to this property, as was first shown in
\cite{MVinstab} and fully analyzed in \cite{MVcomplete}. (The
previously discovered ``gravitational instability'' \cite{NS1}, which
is
also present for the black hole analogues \cite{NS2}, is, however, of a
different
type.) For a study of the problem of
anomalous fermion production in the presence of gravitational solitons,
we refer to the recent paper \cite{MVsthree}.

It is natural to suspect that the existence of the zero-energy solution
found in \cite{gibbons} is mathematically connected with the fact that
the Chern-Simons
number of the Bartnik-McKinnon solution is 1/2, and that one might be
able to generalize the result with the help of an index theorem. In the
present paper we show that this is indeed the case.

In Sections 2 and 3, we prove that the twisted Dirac operator for a
certain
class of gauge fields in a static, regular,
asymptotically flat
space-time has at least two normalizable zero modes, if the
Chern-Simons number of the gauge field is {\em half-integer\/}
(Propositions 1 and 2). The conditions are formulated for any gauge
group and the above statement holds for any representation of the Dirac
spinors (with respect to the gauge group). In Section 4, we show that
this class contains   all CP-symmetric configurations with a connection
that is globally described by a time-independent spatial one-form which
vanishes
sufficiently fast at infinity (Theorem 1).

On the basis of our extensive earlier work on spherically symmetric EYM
solutions for arbitrary gauge groups \cite{OBbirk,OBchern,OBstab}, we
verify in Section 5 that the above-mentioned class contains all
neutral, generic,
spherically symmetric configurations for which the gauge potential is
purely magnetic with real magnetic amplitudes (Theorem 2). [The precise
definition of ``generic'' is given at the
beginning of Section 5.]
In
particular, all neutral, generic solitons of the EYM system belong to
this subclass.

For the benefit of the reader, we remark that the proof of Proposition
2 is mainly based on the Atiyah-Patodi-Singer index theorem for
manifolds with boundary \cite{APS}. However, the verification of the
assumptions of Proposition 2 for
spherically symmetric configurations heavily relies on our
previous work \cite{OBbirk,OBchern,OBstab}, especially on
\cite{OBchern}. In the latter paper, we have also given explicit
expressions for the Chern-Simons numbers of purely magnetic gauge
fields, which involve only general properties of irreducible root
systems and
information about the asymptotic behavior of the configuration.

\setcounter{equation}{0}

\section{Reduction to a harmonic problem on $\mbox{\rm S}^3$}

In a first step, we show for a class of static EYM fields that
normalizable zero modes of the twisted Dirac operator can be obtained
from zero modes of a related Dirac operator on the three-sphere. To
this end, we consider the following set-up:

Let $(M,g)$ be a static
space-time with $M=\bbb{R}\times \S$ and $g=\a^2(dt\,^2-h)$, where
$(\S,h)$ is a three-dimensional Riemannian space and $\a$ is a smooth
function on $\S$. In
addition, let $\cal A$ be a one-form on $\S$ with values in the Lie
algebra $LG$ of the gauge group $G$. We are interested in zero modes of
the
twisted Dirac operator $\bf D$ on $M$ in the background of the one-form
$\cal A$, which is considered as a globally defined gauge potential.

Since the massless Dirac equation is conformally invariant, we may
take
$\a=1$. It is then straightforward to show that $\bf D$ splits in the
Weyl representation of the Dirac-Clifford algebra as follows:
\be
\bdirac=\left(
\begin{array}{cc}
0 & \partial_t\,+\,\mbox{i}\,{\cal D}\\
\partial_t\,-\,\mbox{i}\,{\cal D} & 0
\end{array}
\right)
\;\;,\lb{ns1}
\ee
where $\cal D$ is the three-dimensional Dirac operator on $(\S,h)$
corresponding to $\cal A$ (with one of the two inequivalent
representations of the three-dimensional Clifford algebra).
Hence, the massless Dirac equation,
\be
\mbox{\bf D}\,\Psi=0\;,
\lb{ns2}
\ee
is (for arbitrary $\a$) equivalent to a pair of two-component Weyl
equations,
\be
\mbox{i}\,\partial_t\, \varphi^+={\cal D}\,\varphi^+,
\qquad
\mbox{i}\,\partial_t\,\varphi^-=-\,{\cal D}\,\varphi^-\;.
\lb{ns3}
\ee
Obviously, each (normalizable) zero mode of $\cal D$ provides two
solutions of (\ref{ns2}).

For our applications we now specialize to $\S\approx\bbb{R}^3$ and an
asymptotically flat metric $h$. This property is defined by conformal
compactification: It is assumed that the pull-back $\pi^\ast\,h$ to
$\sthree$ via the stereographic
projection $\pi$ from the north pole can be extended -- after
multiplication with a conformal factor $\Omega^2$ -- to a smooth
Riemannian metric $\tilde{h}$ on $\sthree$. Moreover, the function
$\Omega$ has to vanish with the following rate when the north pole is
approached: If $\Omega$ is expressed in stereographic coordinates
\underline{$x$}, it behaves as $1/|\vec{{$x$}}|^2$ for
$|\vec{{$x$}}|\to\infty$. [Recall that the standard metric on $\sthree$
is conformally Euclidean, with conformal factor
$\O^2=4/(1+|\vec{{$x$}}|^2)^2$.]

If $\cal A$ vanishes at infinity, it is possible to obtain zero
modes of
$\cal D$
on $(\bbb{R}^3,h)$ from zero modes of the Dirac operator
$\tilde{\cdirac}$ on $(\mbox{S}^3,\tilde{h})$ defined by the pull-back
$\tilde{\cal A}$ of $\cal A$.  Clearly, a zero mode $\tilde{\varphi}$
of the Dirac operator $\tilde{\cal{D}}$ gives rise to a normalizable
zero mode of the Dirac operator on $\bbb{R}^3$,
however, with metric $\O^2h$. But now we take advantage of the
well-known
invariance property of the Dirac operator under conformal
transformations to conclude: A zero mode $\tilde{\varphi}$ of
$\tilde{\cal D}$ on $\mbox{S}^3$ gives rise to the zero mode
$\Omega\,\tilde{\varphi}$ of $\cal D$ on $\bbb{R}^3$ equipped with the
{\em original\/} metric $h$. Moreover, since $\tilde{\varphi}$ is
bounded on
$\mbox{S}^3$, the fall-off property of $\Omega$ implies that
$\Omega\,\tilde{\varphi}$ is {\em normalizable\/}.

We thus obtain for each zero mode of $\tilde{\cal D}$ on
$(\mbox{S}^3,\tilde{h})$ two normalizable zero modes of (\ref{ns2}). In
the following sections, we shall investigate for which  EYM
configurations the conditions we had to assume for the gauge fields do
hold.

For reference, we summarize the main conclusion of this section:
\begin{prop}
\lb{p1}
{
Consider a static, regular, asymptotically flat EYM background
configuration with the following property:
\begin{description}
\item[(A1)]There exists a global gauge such that the YM field is
described by time-independent one-form $\cala$ on $\rthree$
which vanishes (sufficiently fast) at infinity.
\end{description}
Let $\tilde{\cala}$ be the pull-back of $\cala$ to $\sthree$ via
stereographic projection\/}, and let $\tilde{h}$ be the metric on
$\sthree$ arising from the background metric by conformal
compactification. Then each zero mode of the twisted Dirac
operator on $(\sthree,\tilde{h})$ belonging to $\tilde{\cala}$ gives
rise to two zero modes of the four-dimensional Dirac operator in the
background of the
given EYM configuration.
\end{prop}

\setcounter{equation}{0}

\section{Zero modes of the twisted Dirac operator on $\sthree$}

By the considerations in section 2, we are led to study the kernel of
the Dirac operator $\tcdirac$ on
$(\sthree,\tilde{h})$ in the background of a global gauge potential
$\tilde{\cala}$ on $\sthree$. (Recall that all principal bundles over
$\sthree$ are trivial, because these are classified by the second
homotopy group of the gauge group $G$.) We are thus interested in
(dropping tildes)
\be
h(\cdirac)=\dim(\mbox{Ker}\,\cdirac)
\lb{ns5}\;.
\ee

In the present section, we prove that the following relation holds for
a certain class of connections:
\be
h(\cdirac)/2 \,+\,\ncs(\cala)\;\in\; \bbb{Z}
\lb{ns6}\;,
\ee
where $\ncs(\cala)$ is the Chern-Simons number of the gauge potential
$\cala$,
\be
\ncs(\cala)=\int_{\;\mbox{$\sthree$}}\;\,\sprod{\cala\,}{d\cala\,+\,
\lbrac{\cala}{\cala}/3\,}
\lb{ns7}\;\;.
\ee
Here, the invariant scalar product in $LG$ is given by
\be
\sprod{X}{Y}=-\frac{1}{\,8\pi^2}\;\mbox{\rm
Tr}\,\{\,L\rho\,(X)\,,\,L\rho\,(Y)\,\}\;\,,
\qquad X,Y\in LG
\lb{ns8}\;,
\ee
where $\rho$ denotes the representation of the Dirac spinors with
respect to the gauge group $G$.

In the following section 4, we will show that all CP-symmetric
configurations satisfy (\ref{ns6}), which, together with Proposition
\ref{p1}, implies our main result: If the Chern-Simons number
$\ncs(\cala)$ of a CP-symmetric gauge potential is {\em
half-integer}, then
$h(\cdirac)$ must be {\em odd\/} and hence, by Proposition \ref{p1},
the
original twisted Dirac operator has at least two normalizable
zero modes.

We establish (\ref{ns6}) with crucial use of the
Atiyah-Patodi-Singer (APS) index theorem for manifolds with boundary,
which we apply to the manifold $I\times\sthree$, where $I$ is an
interval of $\bbb{R}$. Adopting standard notations,
this theorem states that the index of the twisted {\em
Euclidean\/} Dirac operator $\bdirac$ on $M=I\times\sthree$ belonging
to
a gauge potential $A$ on $M$ (and some representation of the gauge
group)
is given by
\be
\ind(\bdirac(A))=-\int_M\hat{A}\wedge\mbox{\rm ch}\;\,-\xi[\partial M]
\lb{ns9}\;,
\ee
where the boundary correction $\xi$,
\be
\xi:=(h+\eta)/2
\lb{ns11}\;,
\ee
is the half-sum of the harmonic correction $h\,[\partial M]$, and
the APS $\eta$-invariant,
\be
\eta[\partial M]\;:=\;\sum_{\{\l_i\neq
0\}}\mbox{sign}\,(\l_i)\;|\l_i|^{-s}\;\;\bigg|_{s=0}
\lb{ns10}\;.
\ee
(For an explanation of the notation, see \cite{eguchi}; for a proof,
see, for instance \cite{APS}.)

In order to apply (\ref{ns9}), we now make the following assumption:
\begin{description}
\item[(A2)]The global gauge potential $\cala$ on $\sthree$ can be
embedded into a one-parameter family $\cala_{(\tau)}$,
$\tau\in\lbrac{-1\,}{1\,}$, with the following properties:
\begin{enumerate}
\item $\cala_{(-1)}$ and  $\cala_{(+1)}$ are pure gauges.\lb{propa1}
\item The corresponding family of Dirac operators
$\cdirac_{(\,\cdot\,)}$ is
``reflection antisymmetric'', that is, $\cdirac_{(+\tau)}$ is conjugate
to \lb{propa2} $-\cdirac_{(-\tau)}$ for all
$\tau\in\lbrac{-1\,}{1\,}$.
\end{enumerate}
\end{description}

An immediate consequence of assumption {\bf (A2.\ref{propa2})} is that
the $\eta$-invariant $\eta_{(\tau)}:=\eta(\cdirac_{(\tau)})$ is {\em
odd\/},
$
\eta_{(+\tau)}=-\eta_{(-\tau)}
$.
Hence,
$
\eta_{(0)}=\eta(\cdirac)=0
$,
which implies that
\be
\xi(\cdirac)=h\,(\cdirac)/2
\lb{ns16}\;.
\ee

Taking advantage of {\bf (A2.\ref{propa1})}, we see that
$
\eta_{(+1)}=\eta_{(-1)}\;,
$
whence
$
\eta_{(\pm 1)}=0
$.
Moreover,
by a theorem due to Lichnerowicz,
$h\,(\cdirac_{(\pm1)})$ = $0$,
since,
up to a gauge transformation,
the Dirac operators $\cdirac_{(\pm
1)}$ agree with the untwisted Dirac operator
$\nabla$. The $\xi$-invariant thus vanishes for $\tau=\pm 1$,
\be
\xi(\cdirac_{(\pm
1)})=0
\lb{ns15}\;.
\ee

Now we are ready to apply the APS index theorem.
The family $\cdirac_{(\,\cdot\,)}$ on $\sthree$ determines the
Euclidean Dirac operator
\be
\bdirac=\mbox{i}\left(
\begin{array}{cc}
0 & \partial_\tau+{\cal D}_{(\,\cdot\,)}\\
\partial_\tau-{\cal D}_{(\,\cdot\,)} & 0
\end{array}
\right)
\lb{ns17}
\ee
on the ``Einstein cylinder'' $\lbrac{-1\,}{1\,}\times\sthree$.
According to (\ref{ns9}), the index of $\bdirac$ on $I\times \sthree$,
where
$I=\lbrac{\tau_1}{\tau_2}\subset\lbrac{-1\,}{1\,}$, is (for inward
oriented normals) given by
\be
\ind(\bdirac)=-\mbox{C}_2[\,I\times\sthree\,]\,-\,\Bigl\{\xi
(\cdirac_{(\tau_1)})+\xi(-\cdirac_{(\tau_2)})\Bigr\}
\;\;,
\lb{ns18}
\ee
where $\mbox{C}_2$ is the second Chern number of the underlying
principal bundle evaluated for the gauge field
$A=\cala_{(\,\cdot\,)}$.
Since $\mbox{C}_2$ agrees with the change $\Delta\ncs$ of the
Chern-Simons number
(\ref{ns7}), and since
$\xi(-\cdirac_{(\tau)})=-\xi(\cdirac_{(\tau)})+h(\cdirac_{(\tau)})$,
Eq.\ (\ref{ns18}) reduces to
\be
\ind(\bdirac)=\Delta\ncs+\Delta\xi-h_{(\tau_2)}
\;.
\lb{ns19}
\ee

The right hand side of Eq.\ (\ref{ns19}) is an integer which only
depends on
the gauge potentials at the boundary. Choosing
$I=\lbrac{-1}{\tau}$, this implies that the following function $f$ is
{\em integer-valued\/}:
\be
f(\tau)\;:=\;\ncs\bigg|_{-1}^{\;\tau}+\;\;\xi\bigg|_{-1}^{\;\tau}-
\;\,h_{(\tau)}\qquad
\in\bbb{Z}\;.
\lb{ns20}
\ee
When evaluated for $\tau=0$, this, together with Eqs.\ (\ref{ns16}) and
(\ref{ns15}), gives
\be
h(\cdirac)/2+\ncs(\cala)\,=\,f(0)+h(\cdirac)+{\ncs}_{(-1)}\;.
\ee

To complete the derivation of relation (\ref{ns6}), it remains to
demonstrate that the Chern-Simons number $\ncs(U^{-1}dU)$ of a
pure gauge $U^{-1}dU$ on $\sthree$ is always an integer. For a proof of
this less surprising statement, we again use the APS index theorem:
Consider the twisted
Dirac operator $\bdirac$ on the Einstein cylinder
$\lbrac{-1\,}{1\,}\times\sthree$
corresponding to the family of gauge potentials defined by
\be
\cala_{(\tau)}=\g(\tau)\,U^{-1}dU\;,
\ee
where $\g$ is a $\bbb{R}_+$-valued function which smoothly interpolates
between zero and one,
\be
\g(-1)=0\;,\qquad \g(1)=1\;,\qquad \mbox{supp}(\dot\g)\subset
(-1\,,\,1)\;.
\ee
For this family, expression (\ref{ns19}) for the index of $\bdirac$
reads
\be
\ind(\bdirac)=\ncs(U^{-1}dU)+\Delta\xi\;,
\ee
since $h_{(1)}$ and ${\ncs}_{(-1)}$ both vanish. Moreover, $\Delta\xi$
also vanishes, since (by construction) the operators $\cdirac_{(-1)}$
and $\cdirac_{(+1)}$ agree up to a gauge transformation. Hence, as
claimed, $\ncs(U^{-1}dU)$ is an integer.

We summarize:
\begin{prop}
\lb{t1}
For a global gauge potential $\tilde{\cala}$ on $(\sthree,\tilde{h})$
with the property {\bf (A2)}, the kernel of the corresponding twisted
Dirac operator $\tilde{\cdirac}$ satisfies
\be
\,\dim(\mbox{\rm Ker}
\,\tilde{\cdirac})/2\;+\;\ncs(\tilde{\cala})\;\in\; \bbb{Z}\;,
\ee
where $\ncs(\tilde{\cala})$ is the {normalized} Chern-Simons number of
the gauge field $\tilde{\cala}$.
\end{prop}

\setcounter{equation}{0}

\section{CP-symmetric gauge fields}

In this section, we establish that the technical assumption {\bf(A2)}
in Proposition
\ref{t1} is always fulfilled, provided that the EYM field is
CP-symmetric. To discuss CP-transformations of gauge fields, we first
show that charge conjugation of a connection form $\omega$ for a
compact gauge group $G$ (which, for simplicity, is assumed to be simply
connected and semi-simple) is induced by the so-called Chevalley
Inversion $\calc$ in the Lie algebra $LG$:
$\omega^{\ctr}=\,\calc\circ\omega$. Hence, for a connection form
$\omega$ on a principal bundle admitting an
automorphism {\rm P} which induces a parity transformation in the base
manifold, there is a natural definition of CP-symmetry: {\em A
connection form $\omega$ is CP-symmetric, if
$\omega^{\cptr}:=\calc\circ \mbox{\rm P}^{\ast}\omega$ is
equal to $\omega$ up to a global gauge transformation\/} (that is, up
to
a bundle automorphism inducing the identity in the base manifold).

\subsection{Chevalley automorphism and charge conjugation of
gauge fields}

The connection between charge conjugation and the Chevalley
automorphism of the Lie algebra $LG$ is most easily discussed in terms
of a (suitably chosen) Chevalley Weyl basis, which is a basis adapted
to a Cartan decomposition of the complexified Lie algebra
$LG_{\compl}$.

For the gauge group $G$, we first fix a maximal torus $T$ and choose a
basis
$S$ of the (real) root system $R$. This basis defines both, the set of
positive
roots $R_+$ as well as the fundamental Weyl chamber $K(S)$,
\be
K(S)=\{\, H\in LT \mid \alpha (H)>0 \,\mbox{ \rm for all } \alpha\in S
\,\} \lb{ns23}\;.
\ee
For the complexified Lie algebra $LG_{\compl}$,
we now choose a
Chevalley-Weyl basis
\be
\{\,h_{\a_j}\,,\,e_\a \mid
\a_j\in  S,\;\a\in R\,\}\;,
\ee
(where $\,h_{\a_j}\in LT_{\compl}$, and $e_\a$ are base vectors of the
root spaces $L_\a$) such that the elements $\mbox{i}h_{\a_j}$ span
$LT$, and such that the set
\be
\{\, \mbox{i}h_{\a_j}\,,\,(e_{+\a}+
e_{-\a})\,,\,\mbox{i}(e_{+\a}- e_{-\a})\mid\a_j\in S,\;\a\in R_+\,\}
\ee
forms a basis of the
compact real form $LG$ (see for instance \cite{Hum}, Sect.\ 25).

Charge conjugation reverses the signs of all charges and thus
corresponds to an
automorphism $\s$ of $LG$ which reduces to $-${\em id} on $LT$. Such an
automorphism is automatically involutive and its extension to the
complexification $LG_{\compl}$ (also denoted by $\s$) interchanges the
root spaces $L_{+\a}$
and $L_{-\a}$.
A particular automorphism with these properties is the above-mentioned
{Chevalley automorphism} $\calc$,
\be
\calc\colon\;\; h_{{\a_j}}\longmapsto -h_{{\a_j}}\;,\quad
e_{+\a}\longmapsto e_{-\a}
\lb{ns24}\;\,.
\ee
Moreover, an {arbitrary} involutive automorphism $\s$ of
$LG_{\compl}$ which leaves $LG$ invariant, and which reduces to $-${\em
id} on $LT$ can be described as follows:
Let $\varphi=\calc\circ\s$, then
\be
\varphi|_{LT}=\mbox{\it id}\;,\qquad \varphi(e_\a)=c_\a\, e_\a\qquad
(\a\in
R)
\lb{ns25}\;\,
\ee
with
\be
|c_\a|^2=1\,,\quad c_{-\a}=\overline{c}_{+\a}\,,\quad c_\a
c_\b=c_{\a+\b}\qquad (\a,\b,\,\a+\b\,\in R)
\lb{ns26}\;\,.
\ee
This implies that the automorphisms $\s$ and $\calc$ are conjugate in
$T\subset G$ (that is, $\s=\Ad(t)\circ\calc$ for some element $t\in T$)
and hence reflects  {a freedom in choosing phases}.

In what
follows, we use the Chevalley automorphism $\calc$ (with respect to a
fixed Chevalley-Weyl basis) to define charge conjugation for a
connection form $\omega$,
\be
\mbox{C}\colon\;\;\omega\longmapsto \omega^{\ctr}:=\,\calc\circ\omega
\lb{ns27}\;\,.
\ee
By the considerations above, any other choice of the automorphism $\s$
would just lead to a compensating gauge transformation with a
{constant} element of the gauge group.

For a given automorphism $\s$, the transformation law for a spinor
field $\Psi$ under C-parity has, of course, to be chosen such that
$\Psi$ and $\bdirac\Psi$ transform in the same manner. Let us add in
this context that the
conjugate representation
$\overline{\rho}$ of a unitary representation $\rho$ is equivalent to
the representation $\rho^{\mbox{\scriptsize\rm C}}:=\,\rho \circ\calc$,
because $\overline{\rho}$ is isomorphic to the
contragradient representation $\rho^\ast$, which has the same weights
as $\rho^{\mbox{\scriptsize\rm C}}$,
$
\overline{\rho}\cong\rho^{\ast}\cong\rho^{\ctr}
$.

\subsection{Existence of zero modes}

Now we are in the position to prove our main result:


\begin{theo}
\lb{theo1}
Under the assumptions of Proposition \ref{p1}, the twisted Dirac
operator
in the background of a CP-symmetric EYM field
has at least two normalizable zero modes, provided that the
normalized Chern-Simons number of the gauge field is half-integer.
\end{theo}

Theorem \ref{theo1} is an immediate consequence of Propositions 1 and
2, as soon as the following assertion is established:

\begin{lemma}
\lb{lalways}
Every CP-symmetric, globally defined gauge field $\tilde{\cala}$ on
$(\sthree,\tilde{h})$ satisfies assumption {\bf (A2)}.
\end{lemma}

For a proof, we first note that CP-transformations are well defined for
the situations we are interested in (because an orientation reversing,
involutive isometry of $(\sthree,\tilde{h})$ with {two} fixed points
can always be lifted to an involutive automorphism of the spin
structure). We also note that $\tilde{\cala}$ (which, by assumption, is
the gauge potential of a CP-symmetric connection) satisfies
\bea
 \tilde{\cala}^{\tcptr}&:=&\,\calc\circ\mbox{\bf p}^\ast\cala
 \nonumber\\
&=&
\Ad(V^{-1})\tilde{\cala}+V^{-1}dV\;,
\lb{cptr}
\eea
where {\bf p} is the parity transformation in the base manifold
$(\sthree,\tilde{h})$, and $V$ is some transition function.

With the help of (\ref{cptr}), we now embed $\tilde{\cala}$ into a
family of gauge potentials with the required properties:
\be
\tilde{\cala}_{(\tau)}:=
\left\{
\begin{array}{crl}
\tilde{\cala}\;+\;\g(\tau)\cdot\Bigl\{\,V^{-1}dV-\tilde{\cala}\,\Bigr\}
\vphantom{\Bigg|}
&\mbox{ for}&\tau\in[\,0\,,1\,]\\
\Ad(V)\tilde{\cala}^{\tcptr}_{(-\tau)} \,+ \,VdV^{-1}&
\mbox{ for}&\tau\in[-1\,,0\,)\;,
\end{array}
\right.
\lb{ns58}
\ee
where $\g\colon[\,0\,,1\,]\to\bbb{R}_+$ is a smooth function
interpolating
between zero and one,
\be
\g(0)=0\;,\qquad \g(1)=1\;,\qquad \mbox{supp}(\dot\g)\subset
(\,0\,,1\,)\;.
\ee

First, we establish {\bf (A2.1)}. Taking advantage of the
transformation
property of a pure gauge $V^{-1}dV$ under a group homomorphism
$\Phi$,
\be
\Phi(V)\,d\,\Phi(V)^{-1}\,=\,L\Phi(V^{-1}dV)\;,
\ee
it is easy to see that
\be
\tilde{\cala}_{(\tau)}=
\left\{
\begin{array}{lrl}
U^{-1}dU&\mbox{ for}&\tau=-1\\
\tilde{\cala}&\mbox{ for}&\tau=\phantom{+}0\\
V^{-1}dV& \mbox{ for}&\tau=\phantom{+}1\;\;,
\end{array}
\right.
\ee
where the pure gauge $U^{-1}dU$ is given by
\be
U^{-1}dU=(V^{\tcptr}V^{-1})^{-1}d(V^{\tcptr}V^{-1})\;.
\ee
Hence, {\bf (A2.1)} is fulfilled.

To deduce the reflection antisymmetry {\bf (A2.2)}, we make use of the
fact that
the twisted Dirac operator
$\tilde{\cdirac}(\tilde{\cala})$ on $(\sthree,\tilde{h})$ transforms
under CP
in the same way as in flat space. (The non-contractibility of $\sthree$
does not induce any serious
global problems.) Up to a gauge transformation, we thus have
\be
\mbox{CP}\circ\tilde{\cdirac}(\tilde{\cala})\;=
\;-\tilde{\cdirac}(\tilde{\cala}^{\tcptr})\circ\mbox{CP}
\lb{ns62}\;.
\ee
This shows {\bf (A2.2)}, since (by construction) the potentials
$\tilde{\cala}_{(\tau)}^{\tcptr}$ and $\tilde{\cala}_{(-\tau)}$ are
gauge equivalent.

\setcounter{equation}{0}

\section{Spherically symmetric EYM fields}
\lb{sec4}

On the basis of our previous work on EYM soliton and black hole
solutions for arbitrary gauge groups \cite{OBbirk,OBchern,OBstab}, we
establish in this section that a ``generic'', spherically symmetric EYM
field (defined in the next paragraph) is CP-symmetric, precisely when
the YM field is purely magnetic with real magnetic amplitudes. As we
shall see, this implies that the remaining technical assumption {\bf
(A1)} in \hbox{Theorem 1} is always fulfilled for fields of this type.

To begin with, we briefly recall the general framework and define what
we call generic EYM fields. A spherically symmetric gauge field is
described by an SU(2)-invariant
connection on a principal bundle $P(M,G)$ admitting an action of the
symmetry group SU(2). In a standard manner, this connection defines
a linear map $\L\colon L\mbox{SU(2)}\to LG$ (the Wang map), which, at
least locally, depends smoothly on the orbits of the symmetry
group in the base manifold (see \cite{OBbirk} or \cite{KN}). The
underlying symmetric bundle $P(M,G)$ is classified by an integral
element
$H_\l$ lying in that part of the integral lattice $I\subset T$ which is
contained in the closed
fundamental Weyl chamber $\overline{K(S)}$, $H_\l\in
I\cap\overline{K(S)}$. In the present discussion, we exclude situations
where this element lies on the boundary
of the fundamental Weyl chamber. The term ``generic'' refers to
configurations for which the classifying element $H_\l$ is contained in
the {\em open\/} Weyl chamber $K(S)$.

In order to avoid extensive repetitions, we adopt the earlier notations
as far as possible. Specific references to formulas in \cite{OBchern}
are indicated by the pre\-fix II.

\subsection{Basic properties of regular configurations}

For a static, spherically symmetric EYM-field, there exists a local
gauge such that the YM field is described by a {time-independent}
potential which, in adapted coordinates, is given by
\be
A=a\,dt+\Lambda (\sigma^{-1}d\sigma)
\lb{ns29}\;.
\ee
Here, $a$ is a $LT$-valued function of the radial coordinate $r$, $\L$
denotes the Wang map, and $\s$ is a local cross section from
$\mbox{S}^2\cong\mbox{SU(2)/U(1)}$ to SU(2), which we choose as
\be
\sigma(\vartheta,\varphi)\,=\,\e^{\;\varphi\,\tau_3}\,\e^{\,\vartheta\,\tau_2}
\lb{ns30}\;,
\ee
where $2\mbox{\rm i}\tau_k=\s_k$ are the Pauli matrices, and
$\vartheta,
\varphi$ denote the standard angular coordinates of $\mbox{S}^2$. For
this section, the potential (\ref{ns29}) takes the form
\be
A\;=\;a\,dt\;+\:{\Lambda_2\,d\vartheta}
+{(\Lambda_3\cos\vartheta-\Lambda_1\sin\vartheta)\,d\varphi}
\lb{ns31}\;,
\ee
where $\L_k:=\L(\tau_k)$ are $LG$-valued functions of the radial
coordinate, satisfying
\be
\L_1=[\L_2,\L_3]\:, \qquad
\L_2=[\L_3,\L_1]\:, \qquad
\L_3 =-H_\l/(4\pi)\:.
\ee
If the configuration is also regular at the origin, the linear map
$L\Phi:=\L|_{r=0}$ is a {\em Lie algebra homomorphism\/} from
$L$SU(2) to $LG$.
Moreover, the same is true for $\L|_{r=\infty}$, provided that the
magnetic charge of the YM field vanishes. (For derivations, see
\cite{OBchern}).

For the discussion of CP-transformations, a very useful technical
feature of regular configurations is that the Wang map $\L$ can be
chosen such that the corresponding homomorphism $\L|_{r=0}$ {\em
commutes} with the Chevalley automorphism. For a proof of this
property, as well as for later use, we first recall that the potential
(\ref{ns29}) is locally gauge equivalent to the following
{\em globally defined\/} gauge potential [Eq.\ (II.44) with a slight
change of notation]:
\be
A=
\Ad\Bigl(\Phi(\s)\Bigr)
\biggl\{\Bigl(\L-L\Phi\Bigr)\,(\s^{-1}d\s)\,+\, a\,dt\,\biggr\}
\lb{ns39}\;,
\ee
where $\Phi$ is the group homomorphism from SU(2) to $G$ corresponding
to the infinitesimal homomorphism
$L\Phi=\L|_{r=0}$. We further recall [see Eq.\ (II.23)] that
$\L_+:=\L_1+\mbox{\rm i}\L_2$ admits an expansion of the form
\be
\L_+\,=\sum_{\a\in S(\l)}\;w_\a\,e_\a
\lb{ns45}\;,
\ee
\be
S(\l)=\{\, \a\in R_+ \mid \,\alpha(H_\l)=2\,\}\;,
\lb{ns34}
\ee
where $e_\a$ are base vectors of the root spaces $L_\a$, {which we
choose to belong to the distinguished Chevalley-Weyl basis introduced
in Section 4.1\/}. Next, we observe that the potential (\ref{ns39}),
after a {constant} gauge transformation with an element $h$ of the
maximal torus $T$, can be rewritten as   \be
A=
\Ad\Bigl((h^{-1}\Phi\,h)(\s)\Bigr)
\biggl\{\Bigl(\Ad(h^{-1})\L-\Ad(h^{-1})L\Phi\Bigr)\,(\s^{-1}d\s)\,+\,
a\,dt\,\biggr\}\;,
\ee
which shows, since
$L(h^{-1}\Phi\,h)=\Ad(h^{-1})L\Phi=\Ad(h^{-1})\L|_{r=0}$, that the Wang
maps $\L$ and $\Ad(h^{-1})\L$, $h\in T$, belong to gauge equivalent
configurations. Using this global gauge freedom, we now fix the phases
of the magnetic amplitudes $w_\a$ such that
the boundary values $w_\a|_{r=0}$ are {\em real\/}. For the
Chevalley transformed ${L\Phi_k}^{\ctr}$, this choice implies
\be
{L\Phi_1}^{\ctr}=-L\Phi_1\,,\qquad {L\Phi_2}^{\ctr}=L\Phi_2\,,\qquad
{L\Phi_3}^{\ctr}=-L\Phi_3\,,
\ee
which proves our claim:
\be
L\Phi(\tau_k)^{\ctr}=L\Phi(\overline{\tau}_k)=L\Phi({\tau_k}^{\ctr})\;.
\lb{comm}
\ee

\subsection{CP-symmetric EYM fields}

For spherically symmetric gauge fields in the global gauge
(\ref{ns39}), we now discuss the action of a CP-transformation, which,
up to a gauge transformation, is given by
\be
\mbox{\bf cp}:A\longmapsto A^{\tcptr}=\,\calc\circ\mbox{\bf p}^\ast
A\;,
\ee
where {\bf p} denotes the parity transformation in the base manifold,
and $\calc$ is the Chevalley-inversion in $LG$.

Since the gauge potential\ (\ref{ns39}) reads more explicitly [see
Eqs.\ (\ref{ns29}) and (\ref{ns31})]
\be
A=
\Ad\Bigl(\Phi(\s)\Bigr)
\biggl\{
\Bigl(\L_2-L\Phi_2\Bigr)d\vartheta\,-\,
\Bigl(\L_1-L\Phi_1\Bigr)\sin\vartheta\,d\varphi
\;+\;a\,dt\,\biggr\}\;\,,
\lb{ns43}
\ee
and since $L\Phi$ is a homomorphism which commutes with the
Chevalley trans\-for\-ma\-tion [see Eq.\ (\ref{comm})], we easily find
\be
A^{\tcptr}=
\Ad\Bigl(\Phi(\s)^{\tcptr}\Bigr)
\biggl\{
-\Bigl(\L_2^{\ctr}-L\Phi_2\Bigr)d\vartheta\nonumber
\,+\,
\Bigl(-\L_1^{\ctr}-L\Phi_1\Bigr)\sin\vartheta\,d\varphi
\;-\;a\,dt\,\biggr\}\;\;,\\
\lb{ns42}
\ee
as well as
\bea
\Ad\Bigl(\Phi(\s)^{\tcptr}\Bigr)
&=& \Ad\Bigl(\Phi(\s^{\tcptr})\Bigr)\nonumber\\
&=&
\Ad\Bigl(\Phi(\,\e^{\,\pi\,\tau_2}\,\s\;\e^{\pm\pi\,\tau_3})\Bigr)
\nonumber\\
&=&
\Ad\Bigl(\e^{\,\pi\,L\Phi_2}\Bigr)
\circ
\Ad\Bigl(\Phi(\s)\Bigr)
\circ
\Ad\Bigl(\e^{\pm\,\pi\,L\Phi_3}\Bigr)\;\,.
\lb{ns42bro}
\eea
To work out the effect of the last map on the right hand side of
Eq.\ (\ref{ns42bro}), we make use of the following general covariance
property of the Wang map $\L$ [see Eq.\ (II.5)]:
\be
\Ad\Bigl(\e^{\;s\,\,L\Phi_3}\Bigr)\circ\L=\L\circ\Ad(\e^{\;s\,\tau_3})\;\,,
\lb{ns40}
\ee
which yields, when evaluated on $\tau_1$ and $\tau_2\,$,
\be
\Ad(\e^{\pm\pi\,L\Phi_3})\,\L_k\;=\;-\L_k\;,\qquad k=1,2
\lb{ns41}\;.
\ee
Hence, apart from a constant gauge
transformation, the action of {\bf cp} amounts to
\be
\mbox{\bf cp}:\;\L_1\longmapsto -\L_1^{\ctr}\;,\qquad
\L_2\longmapsto \phantom{+}\L_2^{\ctr}\;,\qquad
a\longmapsto -a
\lb{ns44}\;\;.
\ee

In terms of the magnetic amplitudes $w_\a$ [see Eq.\ (\ref{ns45})], the
{\bf cp}-\-trans\-for\-ma\-tion (\ref{ns44}) takes the familiar form
\be
\mbox{\bf cp}:\;w_\a\longmapsto \overline{w}_\a
\;,\qquad a\longmapsto -a \;,
\lb{ns46}
\ee
which suggests that for a CP-symmetric configuration all magnetic
amplitudes $w_\a$ are real, and that the electric part $a$ vanishes
identically. In order to demonstrate this, we note that the form of the
potential (\ref{ns39})
remains
unchanged under a gauge transformations with
\be
U=\,\Phi(\s)\,\e^{\,\chi}\,\Phi(\s)^{-1}\;,
\ee
where $\chi$ is a $LT$-valued function independent of the spherical
angles. Since the corresponding transformations of $w_\a$ and $a\,dt$
are given by
\be
w_\a\longmapsto\e^{-\,\mbox{\scriptsize\rm
i}\,\chi_{\a}}\,w_\a\;,\qquad a\,dt\longmapsto
a\,dt+d\chi
\lb{obschrott}
\;,
\ee
where $\chi_{\a}=2\pi\,\a(\chi)$, we conclude that a configuration is
CP-symmetric, precisely when
\be
\overline{w}_\a=\e^{-\,\mbox{\scriptsize\rm i}\,\chi_{\a}}\,w_\a
\;,\qquad a=-d\chi/2
\ee
for some $LT$-valued function $\chi$. This implies $da=a'dr=0$ and
hence $a=0$, since $a|_{r=0}=0$ for regular potentials [see
Eq.\ (\ref{ns39})]. But then the function $\chi$ is constant, which
means that all amplitudes $w_\a$ are real up to constant phases.
Moreover, these phases are trivial, since (by construction) all
boundary values $w_\a|_{r=0}$ are real, and because none of them
vanishes (see \cite{OBchern}, Appendix A). Thus, as claimed,
$\overline{w}_\a=w_\a$ and $a=0$.

Since an arbitrary spherically symmetric metric is CP-symmetric, we
have established the first part of the following intermediate result.
For a proof of the second part, we refer to \cite{OBchern}.
\begin{prop}
\lb{lemma1}
(i) A static, regular, generic EYM configuration is CP-sym\-metric,
if and only if
the YM field is purely magnetic with real magnetic amplitudes.
(ii) For neutral configurations of this type, there exists a
global gauge such that the YM field is described by a time-independent
spatial one-form which vanishes at infinity.
\end{prop}

Proposition \ref{lemma1} classifies the CP-symmetric EYM fields with
spherical symmetry and shows that the remaining technical assumption
{\bf (A1)} in Theorem 1 is always fulfilled for neutral configurations
of this type. As a corollary of Theorem 1, we thus obtain the following
result for spherically symmetric fields:

\begin{theo}
\lb{t2}
For a static, regular, asymptotically flat EYM background configuration
with a YM field that is generic, neutral, and
purely magnetic with real magnetic amplitudes, the corresponding
twisted Dirac operator has at least two
normalizable zero modes, provided that the normalized Chern-Simons
number of the YM field is half-integer.
\end{theo}

For example, all neutral, generic EYM solitons for an arbitrary gauge
group are purely magnetic with real magnetic amplitudes
\cite{OBchern}.
Hence, {Theorem \ref{t2} guarantees the existence of fermion zero modes
for any neutral, generic EYM soliton with a half-integer Chern-Simons
number.\/} (For an explicit verification for SU(2) EYM solitons and
spinors in the fundamental representation see \cite{gibbons}).

\section*{Acknowledgments}
We would like to thank Andreas Wipf for discussions at an earlier stage
of our work. Discussions with members of our theory group, especially
with  Michael Volkov and Markus Heusler, are also gratefully
acknowledged. Finally, we wish to thank the Swiss National Science
Foundation for financial support.


\begin{thebibliography}{50}

\newcommand{\PRL}{{\em \/Phys. Rev. Lett.\/} }
\newcommand{\PR}{{\em \/ Phys. Rev.\/} }
\newcommand{\PL}{{\em \/Phys. Lett.\/} }
\newcommand{\JMP}{{\em \/J. Math. Phys.\/} }
\newcommand{\NP}{{\em \/Nucl. Phys.\/} }
\newcommand{\CMP}{{\em \/Commun. Math. Phys.\/} }
\newcommand{\CQG}{{\em \/Class. Quantum Grav.\/} }



\bibitem{gibbons} G. Gibbons and A. Steif,  \PL {\bf B 314}, 13 (1993)
%
\bibitem{bartnik} R. Bartnik and J. McKinnon, \PRL {\bf 61}, 141 (1988)
%
\bibitem{rubakov} V.\ A.\ Rubakov and M.\ E.\ Shaposhnikov, {\em
\/Phys. Usp.\/} {\bf 39}, 461 (1996); hep-ph/9603208
%
\bibitem{MVsphaleron} D.\ V. Gal'tsov and M.\ S.\ Volkov, \PL {\bf
B273}, 255 (1991)
%
\bibitem{MVinstab} M.\ S.\ Volkov and D.\ V. Gal'tsov, \PL {\bf B341},
279 (1995)
%
\bibitem{MVcomplete} M.\ S.\ Volkov, O.\ Brodbeck, G.\ Lavrelashvili,
and N.\ Straumann, \PL {\bf B349}, 438 (1995)
%
\bibitem{NS1} N. Straumann and Z.-H. Zhou, \PL {\bf B237}, 353 (1990)
%
\bibitem{NS2} N. Straumann and Z.-H. Zhou, \PL {\bf B243}, 33 (1990)
%
\bibitem{MVsthree} M.\ S.\ Volkov, \PR {\bf D54}, 5014 (1996)
%
\bibitem{OBbirk} O. Brodbeck and N. Straumann, \JMP {\bf 34}, 2412
(1993)
%
\bibitem{OBchern} O. Brodbeck and N. Straumann, \JMP {\bf 35}, 899
(1994)
%
\bibitem{OBstab} O. Brodbeck and N. Straumann, \JMP {\bf 37}, 1414
(1996)
%
\bibitem{eguchi} T.\ Eguchi, P.\ B.\ Gilkey, and
A.\ J.\ Hanson; {\em Physics Reports\/} {\bf 66}, 213 (1980)
%
\bibitem{APS} M.\ F.\ Atiyah, V.\ K.\ Patodi, and I.\ M.\ Singer, {\em
\/Math.\ Proc.\ Camb.\ Phil.\ Soc.\/} {\bf 77}, 43 (1975); {\bf 78},
405 (1975); {\bf 79}, 71 (1976);
%
\bibitem{KN} S.\ Kobayashi and K.\ Nomizu {\it Foundations of
Differential Geometry\/} vol 1 (1996), John Wiley \& Sons, New York
%
\bibitem{Hum} J.\ E.\ Humphreys, {\em Introduction to Lie Algebras and
Representation Theory, Graduate Texts in Mathematics\/} {\bf 9} (1972),
Springer-Verlag, New York -- Heidelberg -- Berlin
%
\bibitem{moss} I. Moss and A. Wray, \PR {\bf D46}, 1215 (1992)


\end{thebibliography}
\end{document}